# Recovery of eigenvectors from eigenvalues in systems of coupled harmonic oscillators

Henning U. Voss[*] & Douglas J. Ballon

Citigroup Biomedical Imaging Center and Department of Radiology, Weill Cornell Medicine, New York, NY, USA
*Corresponding author

*Abstract* —The eigenvector-eigenvalue identity relates the eigenvectors of a Hermitian matrix to its eigenvalues and the eigenvalues of its principal submatrices in which the $j$th row and column have been removed. We show that one-dimensional arrays of coupled resonators, described by square matrices with real eigenvalues, provide simple physical systems where this formula can be applied in practice. The subsystems consist of arrays with the $j$th resonator removed, and thus can be realized physically. From their spectra alone, the oscillation modes of the full system can be obtained. This principle of successive single resonator deletions is demonstrated in two experiments of coupled radiofrequency resonator arrays with greater-than-nearest neighbor couplings, in which the spectra are measured with a network analyzer. Both the Hermitian as well as a non-Hermitian case are covered in the experiments. In both cases the experimental eigenvector estimates agree well with numerical simulations if certain consistency conditions imposed by system symmetries are taken into account. In the Hermitian case, these estimates are obtained from resonance spectra alone without knowledge of the system parameters. It remains an interesting problem of physical relevance to find conditions under which the full non-Hermitian eigenvector set can be obtained from the spectra alone.

The eigenvector-eigenvalue identity relates the eigenvectors of Hermitian matrices to their eigenvalues and the eigenvalues of the principal submatrices obtained by removing the $j$th row and column. The recent review by Denton et al. [1] cites Thompson [2] for stating it first in the form used here. Stawiska [3] traced back this identity to Jacobi [4]. Although it has been included in the engineering toolbox for solving inverse problems of vibration for many years [5], it has recently been generating considerable interest [1] due to its application in a theoretical physics problem where it was difficult to compute eigenvectors directly [6]. In graph theory, it has been used to assess the robustness of networks with respect to the removal of network nodes [7]. In general, it may be particularly useful in the study of physical systems where only the spectrum (eigenvalues) of a system described by a Hermitian matrix can be measured, but the modes (eigenvectors) are inaccessible. This situation is often encountered in spectroscopy [8] and is of general interest in quantum systems [9, 10].

We demonstrate that the eigenvector-eigenvalue identity applies to coupled oscillator systems in a way that suggests a simple experimental procedure for recovery of the eigenvectors. The oscillation modes are conventionally obtained as a solution of the eigenvalue problem of the system interaction matrix. If the system matrix is unknown, one can use the eigenvector-eigenvalue identity to obtain the modes from measurements of the system's spectrum and its subspectra by successively removing and replacing each oscillator. This "successive single deletion procedure" will be applied to one-dimensional arrays of coupled radiofrequency (RF) resonators, a system that is easily realized in the laboratory.

The organization of this article is as follows: First, it is shown that coupled oscillator systems are amenable to analysis by the eigenvector-eigenvalue identity. The successive single deletion procedure is then applied to a coupled oscillator model, which is used to estimate the eigenvectors in one-dimensional array experiments. A discussion concludes this article.

The eigenvector-eigenvalue identity relates the $j$th component of the $i$th eigenvector $v$ of a Hermitian $N \times N$ matrix $H$ to its real eigenvalues $\lambda_n(H)$ ($n = 1 \ldots N$) and the real eigenvalues $\lambda_n(H_j)$ ($n = 1 \ldots N$-1) of the principal submatrices $H_j$ obtained by removing the $j$th row and column from $H$. Stated in a form that avoids case distinctions for the occurrence of zero terms [1, 2], it reads

$$|v_{i,j}|^2 \prod_{k=1; k \neq i}^{N} [\lambda_i(H) - \lambda_k(H)] = \prod_{k=1}^{N-1} [\lambda_i(H) - \lambda_k(H_j)] . \tag{1}$$

This identity can be used to obtain the eigenvector components $|v_{i,j}|$ directly from the eigenvalues of $H$ and its principal submatrices $H_j$. These modes are called "Thompson modes" (after Ref. [2]) in the following.

In coupled oscillator arrays, the principal submatrices $H_j$ can be obtained by successively removing and replacing each oscillator in the array. This will be shown in the example of coupled RF oscillators but can be extended to other physical systems, for example mass-spring oscillator arrays [11, 12]. Arrays of $N$ inductively coupled LC resonators with identical geometries, i.e., identical inductances $L_n = L$, are considered here. The mutual inductances between array elements $n$ and $n+1$ are $M_n = \kappa_n L$, where $\kappa_n$ denotes the coupling coefficient between array elements $n$ and $n+1$, and there are $N$-1 couplings. The $\kappa_n$ are dimensionless real numbers and depend on the distances between resonators. The capacitances $C_n$ can be individually set to different values. It is further assumed that the electromagnetic wavelength is large compared to the linear dimensions of the network, such that the near-field regime holds. With the assumption of only nearest-neighbor coupling and the impedance of each isolated element given by $i\omega L + \frac{1}{i\omega C_n}$, the Kirchhoff equations for the currents $I_n$ in the array read

$$\frac{1}{LC_n} I_n = \omega^2 (I_n + \kappa_{n-1} I_{n-1} + \kappa_n I_{n+1}) , \tag{2}$$

with $n = 1 \ldots N$ and fixed-edge boundary conditions $I_0 = I_{N+1} = \kappa_0 = \kappa_N = 0$. For the more general case of non-local couplings, which are studied here, it is advantageous to rewrite the Kirchhoff equations in matrix notation [13, 14] as

$$HI = \lambda I \text{ with } H = CM \text{ and } \lambda = \omega^{-2}. \tag{3}$$

The $N$ oscillator modes each consist of the $N$ current amplitudes $I_n$, and the spectrum consists of the inverse squared resonance frequencies. The capacitance matrix $C$ is diagonal with $C_{nn} = C_n$, and the matrix $M$ contains the magnetic components $L$ and $\kappa_k$ of the system. Its diagonal elements are $L$. Depending on the number of array elements involved in the coupling to element $n$, $M$ has multiple non-zero off-diagonal elements. For example, for nearest neighbor coupling and identical spacing between neighbors, the off-diagonal elements would be $\kappa_1 L$, with a constant coupling coefficient $\kappa_1$. The matrix $M$ for greater-than-nearest-neighbor coupling is given here for the case of $N = 5$ coupled oscillators, which will be used in the experiment below. It is

$$M = L \begin{pmatrix} 1 & \kappa_1 & \kappa_2 & \kappa_3 & \kappa_4 \\ \kappa_1 & 1 & \kappa_1 & \kappa_2 & \kappa_3 \\ \kappa_2 & \kappa_1 & 1 & \kappa_1 & \kappa_2 \\ \kappa_3 & \kappa_2 & \kappa_1 & 1 & \kappa_1 \\ \kappa_4 & \kappa_3 & \kappa_2 & \kappa_1 & 1 \end{pmatrix} . \tag{4}$$

This coupling scheme was chosen since it best approximated the measured couplings of the experimental system as described below. The system matrix is

$$H = CM = L \begin{pmatrix} C_1 & \kappa_1 C_1 & \kappa_2 C_1 & \kappa_3 C_1 & \kappa_4 C_1 \\ \kappa_1 C_2 & C_2 & \kappa_1 C_2 & \kappa_2 C_2 & \kappa_3 C_2 \\ \kappa_2 C_3 & \kappa_1 C_3 & C_3 & \kappa_1 C_3 & \kappa_2 C_3 \\ \kappa_3 C_4 & \kappa_2 C_4 & \kappa_1 C_4 & C_4 & \kappa_1 C_4 \\ \kappa_4 C_5 & \kappa_3 C_5 & \kappa_2 C_5 & \kappa_1 C_5 & C_5 \end{pmatrix}. \tag{5}$$

One observes that unlike $M$ this matrix is not symmetric and thus no longer Hermitian in general. However, since $C$ is diagonal and positive definite, and $M$ is symmetric, the product $CM$ has real eigenvalues [14, 15]. There are two cases to be distinguished, and our two examples to follow match these cases:

1. All capacitances are equal. Then $C$ is a constant diagonal matrix and $H$ is Hermitian. In this case, the eigenvector-eigenvalue identity can be applied without further adjustments.

2. Not all capacitances are equal and $H$ is not Hermitian. For this case we first note that $H = CM$ has the same spectrum as the Hermitian matrix $C^{\frac{1}{2}} M C^{\frac{1}{2}}$. This isospectrality condition [12] follows from

$$\det(\lambda I - CM) = \det\left[ C^{\frac{1}{2}} \left( \lambda I - C^{\frac{1}{2}} M C^{\frac{1}{2}} \right) C^{-\frac{1}{2}} \right] = \det C^{\frac{1}{2}} \det\left( \lambda I - C^{\frac{1}{2}} M C^{\frac{1}{2}} \right) \det C^{-\frac{1}{2}} = 0. \tag{6}$$

In other words, the matrix $C^{\frac{1}{2}} M C^{\frac{1}{2}}$ is symmetric with real eigenvalues, and the eigenvalues of $CM$ and $C^{\frac{1}{2}} M C^{\frac{1}{2}}$ are identical. There are now two systems with identical spectra, $CM$ and $C^{\frac{1}{2}} M C^{\frac{1}{2}}$, but with in general different eigenvectors. Since the eigenvector-eigenvalue identity is designed to find the Hermitian system, an additional step is required to obtain the correct Thompson mode estimates for the non-Hermitian case. Denoting the eigenvector matrix of the Hermitian case with $T$ and the matrix of eigenvalues with $\Lambda$, its eigendecomposition reads

$$T^{-1} C^{\frac{1}{2}} M \, C^{\frac{1}{2}} \, T = \Lambda. \tag{7}$$

The equivalent eigendecomposition for the non-Hermitian case reads

$$U^{-1} CMU = \Lambda. \tag{8}$$

From these two equations follows that the correct (non-normalized) eigenvector matrix $U$ for the system $H = CM$ can be obtained from the transformation [16]

$$U = C^{\frac{1}{2}} T. \tag{9}$$

The Thompson mode matrices corresponding to $U$ and $T$ are the element-wise absolute values of $U$ and $T$, and since $C$ is positive definite, in the last equation $U$ and $T$ can as well be interpreted as the Thompson mode matrices.

In summary, in order to obtain the correct Thompson mode estimates from a system described by the generally non-Hermitian matrix $H = CM$, first the Thompson modes of the similar Hermitian matrix $C^{\frac{1}{2}} M C^{\frac{1}{2}}$ are obtained with the eigenvector-eigenvalue identity, and then Eq. (9) is used to transform those into the Thompson modes of $H$. The price to pay for non-Hermiticity is that the capacitances of the system, or at least their ratios, have to be known a-priori.

Remark: One can define an isospectral family of matrices $C^{\alpha} M C^{1-\alpha}$ ($\alpha \in ]0,1[$). The matrix $CM$ arises from $\alpha \to 1$ and $C^{\frac{1}{2}} M C^{\frac{1}{2}}$ from $\alpha = \frac{1}{2}$. However, the eigenvectors for different $\alpha$ differ in general, and the eigenvector-eigenvalue identity (1) only covers the symmetric case with $\alpha = \frac{1}{2}$.

The procedure to obtain Thompson modes from eigenvalues via the eigenvector-eigenvalue identity has the following correspondence in the RF resonator array model: Removing the $j$th LC resonator amounts to removing the $j$th row and column of both the capacitance matrix $C$ and the magnetic matrix $M$. Fortunately, as $C$ is diagonal, the procedure has the same effect on the product matrix $H = CM$. This is the key requirement for the applicability of the eigenvector-eigenvalue identity to this system. Specifically, removing the second oscillator, for example, the corresponding principal submatrix would be

$$H_2 = L \begin{pmatrix} C_1 & \kappa_2 C_1 & \kappa_3 C_1 & \kappa_4 C_1 \\ \kappa_2 C_3 & C_3 & \kappa_1 C_3 & \kappa_2 C_3 \\ \kappa_3 C_4 & \kappa_1 C_4 & C_4 & \kappa_1 C_4 \\ \kappa_4 C_5 & \kappa_2 C_5 & \kappa_1 C_5 & C_5 \end{pmatrix}. \tag{10}$$

One can see that now the first oscillator couples to the third one (via coupling $\kappa_2$), without involving the second oscillator (coupling $\kappa_1$). Since the distances and thus the couplings themselves do not change, the correct coupling between the first and the third oscillator is indeed given by $\kappa_2$.

Numerical solutions of Eq. (1) for two different arrays with $N = 5$ oscillators were obtained with Matlab (The Mathworks, version R2017a). See Supplemental Material [17] for the code. It also contains specific model parameters such as $L$, $C_1$, $C_2$, and the coupling coefficients $\kappa_1$ to $\kappa_4$. The two arrays are defined as follows:

1. (Hermitian case) An array with identical capacitances $C_1 = C_2 = \ldots = C_5$, which are all defined to correspond to an LC resonator angular base frequency of $f = \omega/2\pi = (LC_1)^{-\frac{1}{2}}/2\pi = 200$ MHz.

2. (Non-Hermitian case) A dimeric array with alternating capacitances, $C_1 = C_3 = C_5$, corresponding to a frequency of $f = 200$ MHz, and $C_2 = C_4$ corresponding to $f = 220$ MHz. Such a dimeric array exhibits a band gap that is evident with five elements but evolves more fully in the asymptotic limit of infinite $N$ [18].

The results are provided in Fig. 1. The spectra of system 1 and 2 are shown in the left panels of Fig. 1A and B, respectively. The bold black lines denote the location of the five principal eigenvalues of $H$. The thin gray lines denote the location of the $5 \times 4$ eigenvalues of the five principal submatrices $H_j$. Due to symmetry, removal of the fourth and fifth resonator duplicates the eigenvalues from the removal of the first and second resonator, and therefore, only $3 \times 4$ distinct subsystem eigenvalues are visible in the spectrum. In the right panels the model modes $|v_{i,j}|$ are displayed in black and, overlaid to them in gray, the Thompson modes obtained via Eq. (1). For the non-Hermitian dimeric system, the additional step of the similarity transformation, Eq. (9), has been applied. In both cases, the model modes and Thompson modes are identical.

In the experiment, the resonator arrays were comprised of $N = 5$ rectangular LC loops that were aligned coaxially and have been described in detail previously [18]. Individual resonators were tuned by a trim capacitor to their base frequencies, which are defined as the resonance frequencies of the isolated, uncoupled, elements. The base frequencies match those used in the numerical simulations of the systems. Thus, the two systems are 1. A system with identical base frequencies of 200 MHz in each of the resonators, and 2. A dimeric system with two alternating base frequencies of 200 and 220 MHz. The spectra of these systems were measured by a network analyzer (Agilent 85046A) with an S-parameter test set. The arrays were inductively coupled to the network analyzer by a loop similar to the LC resonators, without the capacitor. Specifically, the absolute value of the $S_{11}$ input port voltage reflection coefficient versus frequency provided the spectra and subspectra. Spectral peak frequencies were identified by an automatic peak search algorithm. They are provided in the Matlab script.

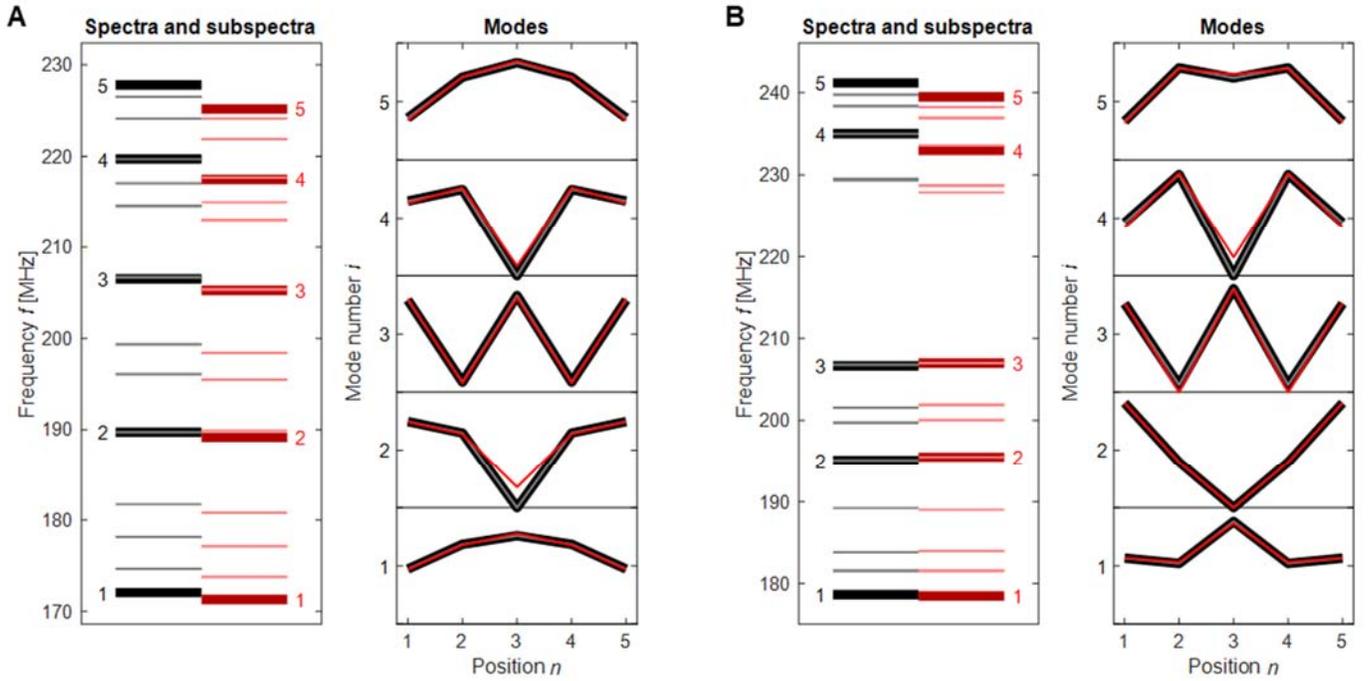

***Figure 1 (color): Obtaining eigenvectors of radiofrequency resonator arrays via the eigenvalue-eigenvector identity.*** *A) Left image: Model spectrum (black) and subspectra (gray), and measured spectrum (dark red) and subspectra (light red) for a system of five coupled identical RF resonators. Right image: Model eigenvector magnitudes (black), model-based Thompson modes (gray), and experimental Thompson modes estimated from the measured spectra shown in the left image (red). B) The same graphs as shown in A) but for a dimeric system with a band gap as explained in the text.*

In order to obtain the Thompson modes from spectra that have unavoidable measurement errors, some adjustments need to be applied to ensure consistency between the measured principal spectrum and the subspectra. Note that the ordered spectrum and subspectra are tightly linked to each other via the Cauchy interlacing inequalities for diagonalizable matrices [1]. These state that the $i$th eigenvalue of each principal submatrix $A_j$ is confined to the interval spanned by the $i$th and $i$+1st eigenvalue of $A$. The Cauchy interlacing inequalities are a necessary condition for $A$ being Hermitian. It has already been shown in Eq. (6) that the characteristic polynomials of the nearly-symmetric matrices $H = CM$, Eq. (3), can be written as characteristic polynomials of real symmetric matrices, such that these inequalities remain valid. However, measurement errors can cause a violation of the Cauchy interlacing inequalities. Measurements also cannot resolve degenerate spectral resonances or resonances that are too close to each other. In the following, a workflow to handle these complications is provided.

*The successive single deletion procedure to obtain modes from experimental spectra:*

1. Measure the spectrum of the system and estimate the spectral peak locations.
2. Remove the first resonator and measure the resulting subspectrum.
3. Replace the first resonator, remove the second one, and proceed until all $N$ subspectra have been measured.
4. Sort the spectra in ascending order.
5. Symmetrize the subspectra according to the symmetry inherent in the array. For example, for the described five-element arrays, removing the first or the fifth, or similarly, the second or the fourth elements should each provide the same subspectrum, but the measurements generally differ. In this case, symmetrization is performed by averaging the first and fifth, and the second and fourth measurements. It turns out that although the measured spectra are already approximately symmetric, this step improves the symmetry of the estimated Thompson modes noticeably.

6. Reconcile the measurements with the Cauchy interlacing inequalities. Measured eigenvalues not adhering to these inequalities are a-posteriori minimally modified to fit into the scheme. For example, if a value is larger than the upper bound specified by the corresponding Cauchy interlacing inequality, it is set to the upper bound. This step helps prevent negative squared mode components. See Supplemental Material [17] for details of this step. It turns out that the necessary adjustments are relatively small.

7. Estimate the Thompson modes from the modified measurements via Eq. (1). In order to identify zero mode components, $|v_{i,j}| = 0$, a numerical threshold is defined that assigns a zero to very small values of the right hand side of Eq. (1). It has been observed that this step contributes to the stability of the procedure including avoiding small negative squared components.

8. For non-Hermitian systems, correct the Thompson modes with Eq. (9).

9. Normalize the estimated Thompson modes in order to compensate for estimation errors.

The spectra of system 1 and 2 are shown in the left panels of Fig. 1A and B, respectively. The bold red lines denote the location of the measured spectral peaks of the resonator array and the thin bright red lines the spectral peak locations of the arrays in which one element has been removed. The right panels contain the experimental Thompson modes in red, overlaid to the black true eigenvector magnitudes and gray model Thompson modes. The experimental Thompson modes match the model Thompson modes. Small deviations can occur for at least two reasons. First, the real system modes are not known but are modeled based on Eq. (3). They depend on the validity of the model equations and accuracy of system parameter estimates that also have an error. Second, errors in the measurements of the eigenvalues propagate into the computation of the Thompson modes via Eq. (1).

It has been demonstrated that the eigenvector-eigenvalue identity is a useful tool to estimate the modes of Hermitian coupled oscillator systems even when the system parameters are not known and the modes cannot be measured directly. This simple formula states that the eigenvalues of a system and specific subsystems encode most of the information about the system dynamics, except phase information. This is remarkable in that explicit knowledge of the system equations besides their symmetry properties is not required. For a special non-Hermitian case, an equivalent solution has been described that involves the a-priori inclusion of some system parameters.

The mathematical procedure of obtaining the necessary principal submatrices of the system matrix translates physically into removal and replacement of single oscillator elements. The procedure only requires the measurement of the system's spectra and subspectra to obtain the system's oscillation modes. This can be of advantage in any physical system where the modes cannot be easily measured otherwise, and simple example systems have been provided. It has been shown that the so derived modes are accurate estimates of model modes if measurements are adjusted in order to obey consistency conditions based on symmetries and the demand that eigenvalues are real. Numerical code for the reproduction of both model simulations and comparison with experimental measurements has been provided.

With only five oscillators, the system considered here is relatively small but allows for resolution of all resonances in the excitation spectra. From the experimental perspective it becomes increasingly difficult to resolve all the spectral peaks as $N$ increases. Typical linewidths of the spectrum of our experimental RF array can be assessed from our previous publication of a similar experiment [18]. From the computational perspective, larger $N$ might cause numerical problems. For example, in order to compute the product terms in the eigenvector-eigenvalue identity it might be prudent to re-arrange the terms in Eq. (6) to avoid numerical overflow of products with many small or large factors. In addition, numerical inaccuracies in the difference terms could cause negative estimates for squared vector components. In this manuscript we have not considered the second degenerate case in the terminology of Denton et al. [1], i.e., eigenvalues of the system that occur with multiplicity larger than one [3]. This could happen in coupled oscillator arrays for example with periodic boundary conditions. Therefore, for situations like these the provided Matlab code needs to be adjusted accordingly.

Equation (1) only defines the magnitude of the eigenvector components but not their signs or phases, which are of importance in interference effects. This is to be expected, as one has the freedom to multiply the eigenvectors by a phase, or, in case of real symmetric matrices, by a sign factor [1]. This reference also discusses how to obtain relative eigenvector component phases for special cases. The feasibility of these methods in specific physical applications remains to be investigated.

Potential applications of the successive single deletion procedure to obtain oscillation modes from measured spectra include magnetic resonance imaging (MRI). Arrays of coupled radiofrequency resonator elements are used routinely in MRI [11, 14, 19-24]. It is well known that loading of the array with a human subject alters the eigenvalues in the dispersion relation in sometimes unpredictable ways. As a consequence, the eigenmodes, from which the RF field distribution is calculated via the Biot-Savart law, are also modified. In MRI, a single mode is typically chosen for imaging. Using the above methodology, the imaging mode could be calculated in the loaded array by first measuring the full spectrum and then successively removing array elements by opening or shorting the loops remotely to obtain the subspectra. It would then be possible to iteratively adjust the capacitances $C_n$ on the arrays to converge to the most uniform RF field distribution possible. This procedure may assist in the design of resonator arrays for example for ultra-high field MRI.

The eigenvector-eigenvalue identity can be seen as a map from the eigenvalue spectrum of a Hermitian matrix to its set of eigenvectors. The system matrix of the monomeric system considered here is in fact Hermitian and its eigenmodes could be straightforwardly estimated from measurements. The matrix of the dimeric system is non-Hermitian, but the dimeric system is in fact more interesting as it can give rise to localized edge modes [18]. This matrix has the same eigenvalue spectrum as a similar Hermitian matrix [16], and a naïve application of the eigenvector-eigenvalue identity to its spectrum naturally provides the eigenvectors of the similar Hermitian matrix. The manner in which the eigenvectors of a non-Hermitian matrix can be recovered alone from its spectra via the eigenvector-eigenvalue identity and some minimal additional assumptions remains an interesting problem.

Disclosure: The authors and their institution, Cornell University, own patents that are related to radiofrequency resonators similar to the ones being used in this manuscript.

---